\documentclass[12pt,a4paper]{article}
\newcommand{\be}{\begin{equation}}
\newcommand{\ee}{\end{equation}}
\newcommand{\bea}{\begin{eqnarray}}
\newcommand{\eea}{\end{eqnarray}}
\newcommand{\beas}{\begin{eqnarray*}}
\newcommand{\eeas}{\end{eqnarray*}}
\newcommand{\ba}{\begin{array}}
\newcommand{\ea}{\end{array}}
\newcommand{\nn}{\nonumber}

\newcommand{\bt}{\begin{table}}

\newcommand{\vsi}{\varsigma}

\newcommand{\ga}{\gamma}

\newcommand{\ka}{\kappa}
\newcommand{\la}{\lambda}

\newcommand{\si}{\sigma} 
\newcommand{\Si}{\Sigma}

\begin{document}

\title{\bf Unimodular bimode gravity, grand unification and the
scalar-graviton\\ dark matter}
\author{Yu.~F.~Pirogov
\\
\small{\em Theory Division, Institute for High Energy Physics,  Protvino, 
Moscow Region, Russia }
}
\date{}
\maketitle

\begin{abstract}
\noindent
In prior  article of the author,
the unimodular bimode gravity/{\em systo-gravity}, with the
scalar-graviton/{\em systolon}  dark matter, was worked out. 
To compile with the anomalous  rotation curves of galaxies the scale of the
local scale violation in the theory was shown to be about $10^{15}$~GeV.
In this letter,  to naturally incorporate such a  scale the
{\em hyper unification} framework,  merging  systo-gravity
with  grand unification  through matter, is constructed.
The systolon, as a  free propagating compression mode   in metric,   emerges
only  below the unification scale,  possessing  at the same time a modified
high-energy behaviour to be manifested at the high temperatures.\\

\end{abstract}

\section{Introduction}

Conceivably,   General Relativity (GR) is not the fundamental theory of
gravity being rather  a  (principle)  part of the effective field theory
of metric, which  may a priori comprise other  propagating   gravity modes in
addition to the (massless transverse-tensor) graviton.
In this vein,  in looking for  a gravitational  dark matter (DM) interior to
metric, the unimodular bimode gravity (UBG) was proposed~\cite{UBG}.
UBG expli\-citly violates the general gauge invariance/relativity to the
residual unimodular one remaining still general covariant.
Due to violation of the local scale invariance there emerges in the
metric  a (light) scalar graviton/systolon treated as   DM. It proves that
in order to compile with the  galaxy dark halos   the scale of violation of the
local scale invariance  should  lie in  the grand unified theory (GUT)  range,
about $10^{15}$~GeV. So the question arises  whether  this could  be more
then just a coincidence? This letter is an extension to~\cite{UBG}. Here we try
to answer the posed question in affirmative by embedding
UBG into the GUT  framework. To make the exposition
self-contained  as far as possible,  we   first recapitulate the essence of
UBG, necessary for  the following,\footnote{For more
detail, see~\cite{UBG}.}   mer\-ging then UBG and  GUT, with the natural
emergence of the required scale.

\section{UBG/systo-gravity}

Comprising   the systolon and graviton, UBG will   otherwise  be referred
to as the {\em systo-gravity}. The proper Lagrangian looks generically like
\be
L_{sg }= L_g+L_s, 
\ee
with contributions of  graviton, $L_g$, and systolon, $L_s$, respectively,
preserving and violating  the general
invariance/relativity. As a para\-digm, choose for  $L_g$ the
GR Lagrangian 
\begin{equation}
\label{GR}
L_{g}=- \frac{1}{2}  \kappa_g^2R,
\end{equation}
where  $R $ is the Ricci scalar for metric $g_{\mu\nu}$. Here one puts
$\ka_g=1/\sqrt{8\pi G} =1.4 \times 10^{18}$ GeV, with   $ G$ being  the Newton
constant. Some  general invariant modification of $L_g$, say, through  a
function $f(R)$ is a priori admitted. 

Let  now the general 
invariance/rela\-tivity be  vio\-lated to the residual unimodular one, 
$G\to U$, by means  of the explicit dependence of $L_s$ on the scalar density
$g=\det g_{\mu\nu}$. To maintain  the general covariance $g$ should enter
through $g/ g_\ast$, where   $g_\ast$ is a non-dynamical scalar
density of the same weight as the dynamical $g$.
More particularly, without loss of generality represent  systolon  by   the
unimodular invariant and general
covariant (dimensionless) scalar field
\be\label{Sg}
\vsi=\ln \sqrt{-g}/ \sqrt{-g_\ast}.
\ee
A priori, $L_s$  is an arbitrary function of $\vsi$.
To  terminate the arbitrariness let us  enhance $U$  by a dynamical global
symmetry defined in any fixed coordinates $x^\mu$ (with no coordinate change)
through  the field substitutions:
\bea\label{gt}
g_{\mu\nu}(x)& \to& \bar g_{\mu\nu}(x)=e^{-2\la_0}
g_{\mu\nu}  (e^{-\la_0} x),\nn\\
g(x)&\to &\bar g(x) =e^{-8\la_0}g(e^{-\la_0} x )   ,
\eea
with an arbitrary  parameter   $\la_0$, whereas
\be \label{g2}
g_\ast(x)\to  \bar g_\ast(x) = g_\ast (e^{-\la_0}x ) .
\ee
The symmetry  is aimed at  distinguishing the  
non-dynami\-cal and dynamical    fields. For the latter ones    these
transformations   (followed by the coordinate substitutions  $x^\mu\to \bar
x^\mu=e^{\la_0} x^\mu$) coincide  with the global scale transformations  as a
part of the general coordinate transformations. For this reason  the general
invariant part of $L_{sg}$, $L_g$, is also global symmetric.  
Eqs.~(\ref{gt}) and (\ref{g2})
may be referred to as the compression transformations. In these terms, 
the non-dynamical measure  $\sqrt{-g_\ast}$ is  ``incompressible'' in contrast
to the dynamical  one $\sqrt{-g} $. 
It follows  thereby  that   $\vsi$ transforms
inhomogeneously under compressions:
\be\label{gs}
\vsi(x)\to  \bar \vsi(x)= \vsi(e^{-\la_0 } x)  -4 \la_0 .
\ee
In the so-called ``transverse''
coordinates/gauge, where $g_\ast$ $ = \bar g_\ast =-1$,  compressions are
equivalent to the global scale transformations/di\-latations.\footnote{ Under 
the (unbroken) compression symmetry,  the systolon may be considered as a
general covariant counterpart of  dilaton, the latter being
the Goldstone boson corresponding  to the (non-linearly realized broken)
dilatation symmetry. Clearly, their  underlying  motivations and meanings  are 
different.} 

Such a  (approximate) global  symmetry  may
serve as a {\em raison d'\^{e}tre} for suppression of   the derivativeless  
coup\-lings of systolon. 
In terms of  the dimensionfull  scalar field $\si\equiv
\kappa_s\vsi$,  the admitted Lagrangian looks  like
\be\label{s} 
L_s=\frac {1}{2} g^{\mu\nu}\partial_\mu \si  \partial_\nu \si +V_s(\si),
\ee
where  $\ka_s\leq \ka_g$ (or rather $\ka_s/ \ka_g$)  is  a free parameter of the
scalar gravi\-ty, treated as  the scale of the local scale violation. 
The higher-derivative terms are to be
suppressed by the powers of  $\ka_s$ (or, possibly, $\ka_g$).
For generality  we still explicitly  retain  the potential~$V_s$ (whose 
quadratic part gives a  mass to systolon),  
assuming it to be   the leading correction to the otherwise  global  symmetric 
gravity Lagrangian. 
The (light) systolon presents a scalar compression mode in metric in addition to
the transverse-tensor, four-volume preserving deformation  mo\-de presented  by
the (massless) graviton.
The systo-gravity,  due to   constraint (\ref{Sg}) and the emergent
spontaneous breaking of the compression symmetry,  is apt to describe
DM in the galaxy dark halos. To this end $\ka_s/\ka_g$ ought  to be of
order  $v_\infty/c\sim 10^{-3}$, where $v_\infty$ is the asymptotic rotation
velocity  in galaxies, i.e., $\ka_s$ should lie in the GUT range, about
$10^{15}$ GeV. Thus merging systo-gravity with GUT is natural. 

\section{Grand/hyper  unification}

At face value, the theory presented  so far concerns  the   pure gravity. To
account for matter let us address ourselves to  a  generic (renormalizable) GUT
taking its general invariant Lagrangian  as the matter one:
\be\label{Lm}
L_{m} = L_{m}(g_{\mu\nu}, V_\mu, \Phi,F). 
\ee
Here  $V_\mu$, $\Phi$ and $F$ are the
generic gauge, scalar and chiral fermion fields, respectively.
In particular,  (\ref{Lm}) includes a  scalar potential $V_m(\Phi)$, whose
(absolute) minimum defines the vacuum expectation values   $\Phi_0$'s of
$\Phi$'s. The detailed expressions are  of no principle
importance  for what follows.
Now  switch on the   systolon-matter interactions violating general
relativity,  with the unsuppressed  unimodular invariant effective 
Lagrangian (of the canoni\-cal dimension $d=4$)  as follows:
\bea\label{sm}
L_{sm}&=& \frac{1}{2}  (\Si \xi_\Phi \Phi^\dagger \Phi
)g^{\mu\nu}\partial_\mu
\vsi
\partial_\nu \vsi\nn\\
&+&g^{\mu\nu}(i\Si \xi'_\Phi \Phi^\dagger     \partial_\mu
\Phi+\mbox{\rm h.c.})\partial_\nu \vsi\nn\\
&+&g^{\mu\nu}(\Si \xi_F \bar F \ga_\mu F)\partial_\nu \vsi .
\eea
Here  $\xi_\Phi$, $\xi'_\Phi$ and  $\xi_F$ are some dimensionless parameters 
presumably of order unity. 
The gauge fields, which  enter through the square of the covariant  gauge
strength $V_{\mu\nu}$ of the canonical dimension $d=2$, do not admit a similar
term.   From the point of view of GUT alone, (\ref{sm}) is a priori  as good as
(\ref{Lm}). In the case of absence  of any other dimensionfull 
parameter, but for $\ka_g$, all the
higher-dimension terms ought to be suppressed by  powers of~$\ka_g$.

After the spontaneous symmetry breaking, $\Phi= \Phi_0+  \Delta\Phi$, 
with $\Delta\Phi$'s being the physical scalar fields, one~gets
\bea\label{sm'}
L_{sm}&= & \frac {1}{2}\Big(1+\frac{1}{ \ka_s^2} \Delta
N(\Phi_0,\Delta\Phi)\Big)
 g^{\mu\nu}\partial_\mu \si \partial_\nu \si \nn\\
&+&    \frac{1}{ \ka_s} g^{\mu\nu} 
J_\mu(\Phi_0,\Delta\Phi,F)\partial_\nu \si,
\eea
where $J_\mu$ and  $\Delta N$  are, respectively, the ensuing  matter current
and the  correction to the  kinetic term normalization. At that, $\ka_s$ gets
not independent:\footnote{Particularly,    $\ka_s$ should depend  on the state
of the matter vacuum, e.g.,  its   temperature.}
\be
\kappa_s^2=\Si  \xi_\Phi
\Phi_0^\dagger \Phi_0.
\ee
Now we can  completely abandon the kinetic term for $\si$
in (\ref{s}) in favour of (\ref{sm'}). 
In this approach  the pure gravity is general
invariant (as in GR or in its general invariant modifications).
Merging systo-gravity with grand
unification may be referred to as the {\em hyper unification}.
Above   the  scale $\ka_s$   the systolon, as a
free propaga\-ting particle, disappears.
For   the finite field $\si$,  its  residual interactions  with  
light particles  in  (\ref{sm'})  are naturally suppressed  at least as
${\cal O}(1/\ka_s)$. The emergent decoupling of the systolon DM from the
ordinary  matter may serve as an additional reason in favour of the proposed
hyper unification.

\section{Conclusion}

The inclusion of the local scale violation   into
the theory of particle interactions  via   GUT's
seems  theoretically very attractive as relating the drastically different 
distances   in nature.   If relevant, the hyper unification incorporating  grand
unification   and systo-gravity, with  systolon DM,  may stretch  down from the
Planck scale  to the cosmological distances. The phenomenological verification
of the theory  in  cosmology,  both at the   large
distances, notably with respect to the systolon DM,   and  at the high
temperatures, would be crucial.\footnote{Under 
the hyper unification,   the  two-component DM, with
a  particle admixture  to the systolon  DM, is feasible. One would  though
expect  the particle DM to be  relevant (if any) near the core of dark halos,
while the coherent systolon field off the core.}

\end{document}